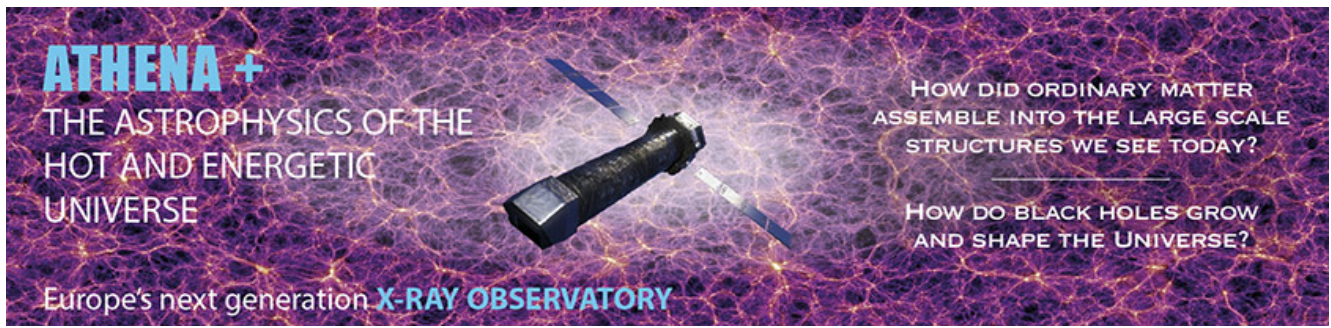

# The Hot and Energetic Universe

An *Athena+* supporting paper

## The close environments of supermassive black holes

### Authors and contributors

**Michal Dovciak, Giorgio Matt,** Stefano Bianchi, Thomas Boller, Laura Brenneman, Michal Bursa, Antonino D'Ai, Tiziana di Salvo, Barbara de Marco, Rene Goosmann, Vladimir Karas, Kazushi Iwasawa, Erin Kara, Jon Miller, Giovanni Miniutti, Iossif Papadakis, Pierre-Olivier Petrucci, Gabriele Ponti, Delphine Porquet, Chris Reynolds, Guido Risaliti, Agata Rozanska, Luca Zampieri, Andreas Zezas, Andrew Young

# 1. EXECUTIVE SUMMARY

Most of the action in an AGN occurs in a very small region, within a few tens of gravitational radii from the supermassive black hole. Within this region, matter in the accretion disk may lose up to almost half of its energy, resulting in a copious production of X-rays. It is thought that winds and jets are launched from this area, and this region is also where the effects of General and Special relativity most prominently leave their mark on the emitted radiation. Consequently, the immediate vicinity of the event horizon is where differences in the metric of the spacetime due to the black hole's rotation become appreciable. Understanding what happens in the close environment of the black hole yields critical insight into four open areas of research in astrophysics: (1) how the most efficient energy release mechanism in the Universe works, (2) how strong gravity affects the behaviour of matter and radiation, (3) how the physics of accretion and ejection are related, and (4) whether or not black holes are spinning, which in turn tells us about the history of their growth.

X-rays in AGN are produced by Comptonisation of thermal disk photons in a hot corona (see Fig. 1). Part of the resulting power-law continuum illuminates the disk, where it is reprocessed and reflected, both in soft and hard X-rays. However, we still do not have an answer to some crucial questions: does the disk always extend down to the innermost stable circular orbit (ISCO) or is it truncated under certain conditions (which may be related to the launch of winds and jets)? What is the location and geometry of the hot corona? Is it compact (as would be the case if the corona actually represents the base of a jet) or rather clumpy and spread over the disk (as would be the case if it is produced by local disk instabilities, probably of magnetic origin)? Mapping the disk-corona geometry is therefore the key to understanding where the energy powering the X-ray emitting corona ultimately comes from (e.g. the disk or the rotating black hole). Observations of time lags between the reprocessed emission from the disk and the primary radiation from the corona provide the most promising tool to answer these questions. Time lags measure the light-crossing time between these two regions, and can therefore be used to measure the distance between them. This is definitely the most promising method to map the inner regions of AGN. XMM-Newton discovered time lags in several AGN, but a full exploitation of this technique — including the ultimate goal, the determination of the disk-corona geometry via transfer function fitting — requires data of much better quality. Only *Athena+* can provide such data, with its very large effective area over a broad range of energies, the low background ensured by its focusing optics, and the possibility of long uninterrupted observations on its orbit at L2.

The spectrum of the reprocessed emission is strongly modified by both General and Special Relativistic effects, and can be fitted with models to derive the main disk parameters, including the inner radius of the disk. Under the assumption that it coincides with the ISCO, and using the ISCO dependence on the spin of the black hole, the latter parameter — which is crucial to understanding the accretion history of supermassive black holes — may be determined. In this respect, the improvement gained with *Athena+* is two-fold. The large increase in effective area over a broad energy range will enable sensitive measurements of all the reflection features (lines and continuum), and will allow us to measure the spins of black holes beyond the local Universe. Moreover, the excellent energy resolution of *Athena+* will enable emission line components originating in more distant matter (ubiquitously present in the X-ray spectra of AGN) to be separated from broad emission features originating in the innermost disk.

Copious amounts of intervening gas are frequently present within AGN along our line of sight at larger distances from the black hole, resulting in additional spectral signatures superposed on the X-ray continuum and emission features. Orbiting clouds and outflowing winds are likely present up to a few light days from the black hole, in the so-called Broad Line Region, while at scales of 0.1–1 pc a low ionization, dusty torus-like structure is present. Primary X-rays are reprocessed in all of these regions, and strong absorption and emission features are produced. The combination of the large effective area and fine energy resolution of *Athena+* will enable the various spectral components to be deconvolved (see also Cappi, Done, et al., 2013, *Athena+* supporting paper). The distance of these regions from the black hole can be estimated by measuring the line widths, which also yields an estimate of the amount of matter residing in each reservoir of gas. Inflow/outflow velocities of the gas can also be measured using this technique, providing a link between the close environment of the SMBH and its host galaxy.

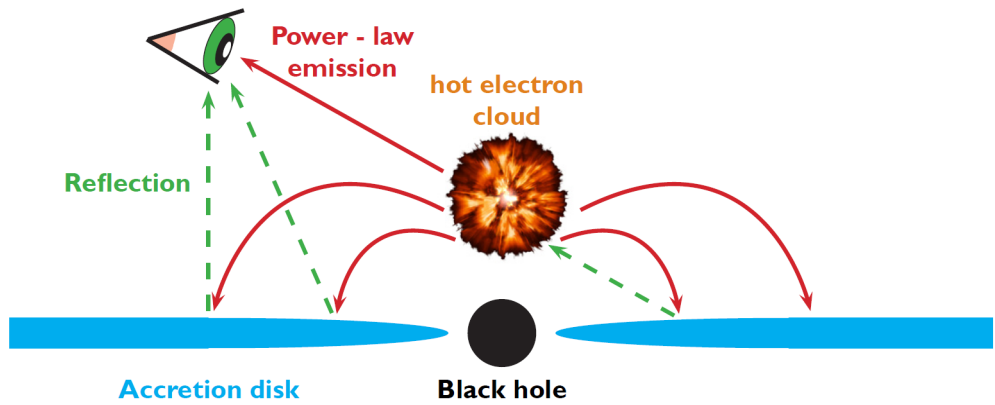

Figure 1: Cartoon of a black hole and accretion disk (blue) above which the power-law continuum is generated by Comptonisation of thermal disk photons in a hot electron cloud (orange). Power-law emission (red) irradiates the disk producing the reflection spectrum (green). Variations in the power-law emission create delayed, blurred changes (reverberation) in reflection. Thanks to *Athena+* it will be possible to translate these signals into a physical picture of the inner region around the black hole.

## 2. EXPLORING THE CLOSE ENVIRONMENT OF THE BLACK HOLE

As one approaches the black hole, strong gravity effects (which are a combination of General and Special relativistic effects) become more and more important and shape the emission from this region in a very distinctive way, providing a number of observational indicators that allow us to map the location and motion of the matter therein. When radiation is emitted at a few gravitational radii, these effects must be treated in full GR (weak-field approximations are no longer valid) offering us the opportunity to test GR against competing gravity theories.

### 2.1. Time lags and reverberation mapping

Measuring the light travel time between flux variations in the hard X-ray continuum and the lines that it excites in the accretion disk (light echo or "reverberation"; e.g. Stella 1990, Matt & Perola 1992) provides a model-independent way to map the inner accretion flow, as the time delay simply translates into distance for a given geometry (after proper relativistic corrections are applied). Reflection of the continuum in hard and, if matter is at least partly ionized, also in soft X-rays occurs along with the line emission (Ross & Fabian 2005). Soft X-ray reverberation lags were indeed seen with XMM-Newton in the Narrow Line Seyfert 1 (NLS1) galaxy 1H0707-495 (Fabian et al. 2009), where a reverberation lag of 30s at high frequencies has been detected. Similar lags were later discovered in other bright NLS1s (Emmanoulopoulos et al. 2011), but also in normal Seyferts, where they may be relatively common, as suggested by the work of De Marco et al. (2013) who also reported evidence for a correlation between the lag time scale and the mass of the black holes. These lags are fully consistent with, and strongly support, the reflection scenario. The observed time lags clearly indicate that these regions are within at most a few tens of gravitational radii from the black hole (e.g. Reis & Miller 2013), and therefore that X-rays really probe the innermost regions of the AGN. The discovery that these lags are more pronounced at the iron line energy (Zoghbi et al. 2012, 2013) further supports this interpretation.

Results from XMM-Newton are however restricted to a relatively small number of AGN and a relatively narrow frequency range. The much larger effective area of *Athena+* will allow us to estimate the time lags spectrum over a broader frequency range, and with a significantly improved precision (see Figs. 2 and 3). These are necessary in order to extend these studies to many more sources and to determine accurately the "transfer function" between the continuum and the reprocessed emission (Blandford & McKee 1982, Wilkins & Fabian 2013), and consequently map precisely both the emitting and reflecting regions. This is the best technique to determine the geometry of the innermost region and the disk-corona system, which is fundamental to understand how AGN really work.

See animations at http://stronggravity.eu/public-outreach/animations/reverberation for graphic explanations of these phenomena.

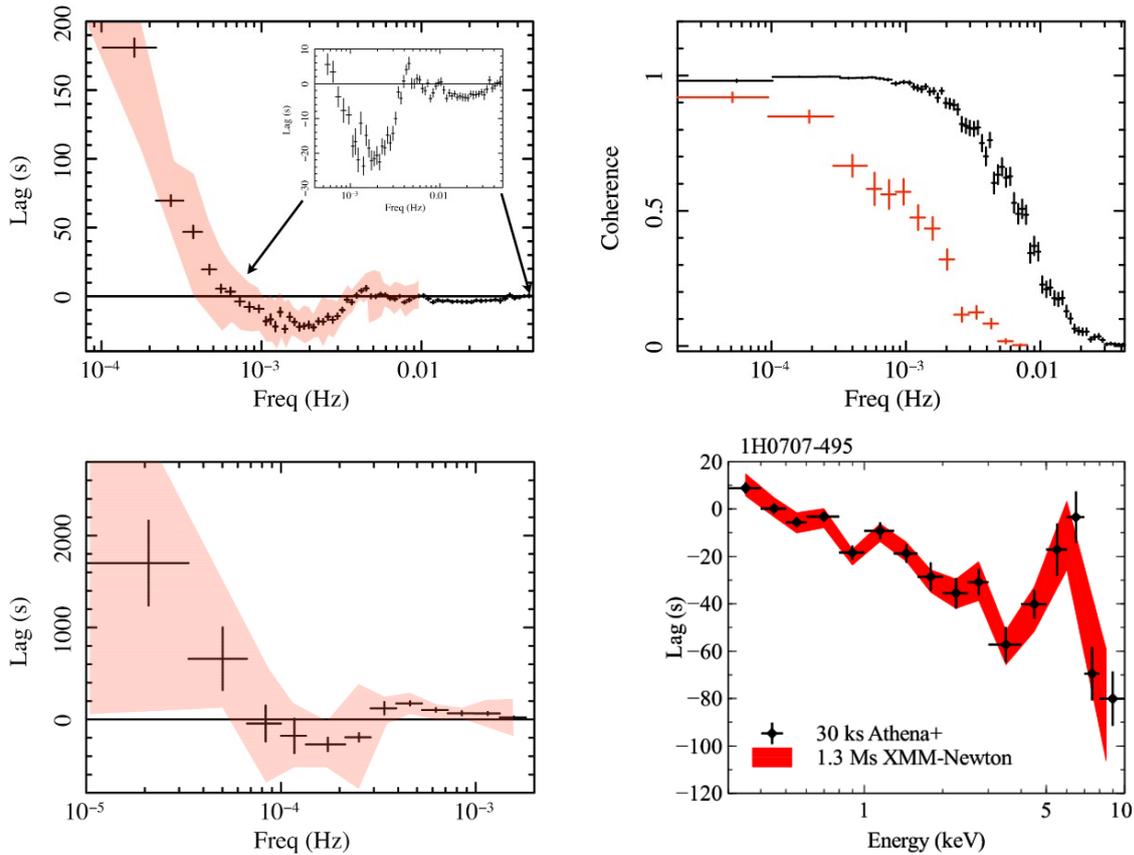

Figure 2: *Left upper panel:* expected time lags (1–4 keV against 0.3–1 keV) as a function of frequency with *Athena+* for 1H0707-495 (exposure time as in the XMM observation, i.e. 500 ks). The red region is the 1σ contour with XMM. While positive lags below 0.001 Hz are likely due to perturbations in the disk, negative lags at high frequencies (i.e. small time scales) measure the light-crossing time required by the primary X-ray photons to reach the disk, and therefore the average distance between the two regions. Note that structures at frequencies larger than 0.01 Hz, which are inaccessible with XMM-Newton, can also be studied. *Right upper panel:* the raw coherence function for XMM (red) and *Athena+* (black). The drop due to Poisson noise is shifted to higher frequencies, which allows detection of lags in a much broader frequency range. *Left lower panel:* the expected time lags vs. frequency with *Athena+* for the Seyfert galaxy IC4329A, using the XMM parameters as inputs. In XMM the detection was not significant (see the red region, representing the XMM 1σ contour). *Right lower panel:* the expected time lag vs. energy for 1H0707-495 with *Athena+* 30 ks short observation compared to 1.3Ms long XMM observation. The full line profile will clearly emerge in the observed energy dependence of lags. The lag spectrum will then complement the photon spectrum allowing for an unambiguous decomposition into the various emitting components.

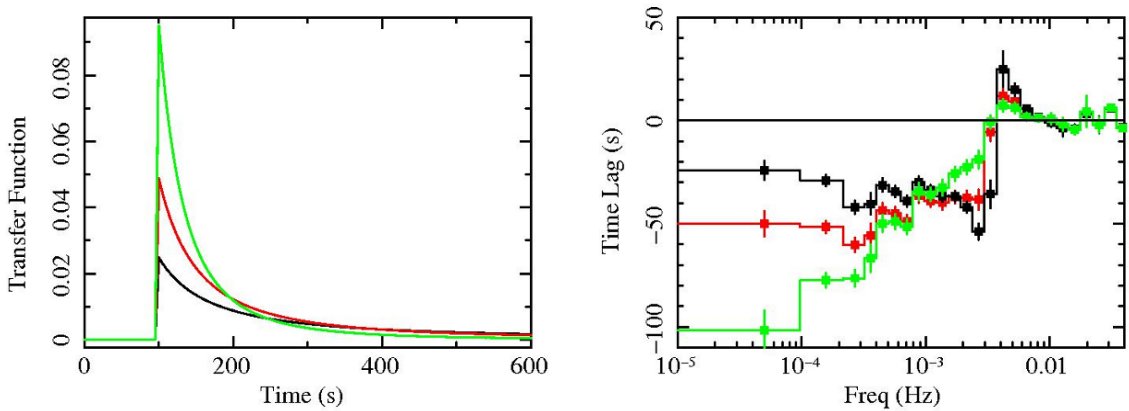

Figure 3: *Left:* a set of transfer functions used to simulate the time lags shown in the right panel. Each function corresponds to a different geometry of the emitting corona. The sharper the tail, the more compact the illuminating region. *Right:* the resulting time lags vs. frequency (disk perturbations dominating at low frequencies are not simulated). *Athena+* has the ability to distinguish between different transfer functions, and hence different geometries of the disc-corona system.

## 2.2. Measuring the spins of black holes

Black holes are characterized by only two parameters, their mass and their spin. A proper census of the black holes in the Universe therefore requires the measurements of both parameters. Measuring the spins of the SMBHs in a sizeable number of AGN is not only interesting per se, it is also of fundamental importance to understanding the black hole growth history in the local Universe (Berti & Volonteri 2008), particularly the relative roles of mergers vs. prolonged and chaotic gas accretion (Fig. 5). However, while the mass influences its environment up to relatively large radii (the sphere of influence is of the order of a few parsecs for a black hole of 10 million solar masses), the influence of the spin is felt only up to a few gravitational radii. Therefore, measurements of black hole spin in the electromagnetic domain can be best done in X-rays. The simplest and most widely applicable way of measuring BH spin is via time-averaged spectral fitting of relativistic iron Kα lines (Fabian et al. 2000), which are present in at least ~30–40% of all AGN (Nandra et al. 2007, De la Calle Perez et al. 2010). The prevalence of these lines demonstrates that they are a natural consequence of accretion onto compact objects, and estimates of BH spin have already been made in a few AGN (e.g. Brenneman & Reynolds 2006, Brenneman et al. 2011, Risaliti et al. 2013, Walton et al. 2013). As already mentioned, the iron line is always accompanied by a reflection continuum in hard X-rays and, if the matter is at least partly ionized, also in the soft X-rays. With its large effective area over a broad energy range, *Athena+* will permit the simultaneous use of the iron line and the soft X-ray reflection continuum to measure black hole spins. Moreover, its excellent energy resolution will easily separate the broad from the narrow components, which are ubiquitous in AGN and originate from more distant matter (Yaqoob & Padmanabhan 2004; see Fig. 4).

The very large effective area of *Athena+* will also allow measurements of black hole spins in sources well beyond the local Universe. As an example, a maximally rotating black hole spin in PB5062, a luminous ($L_x \sim 3 \times 10^{46}$ cgs) QSO at $z=1.77$ (from the CAIXA catalog, Bianchi et al. 2009) can be recovered with a precision of 20% in a 100 ks observation. Even if the number of sources for which such measurements can be performed is necessarily limited, it will be possible to put interesting constraints on the spins of the black holes at intermediate redshifts.

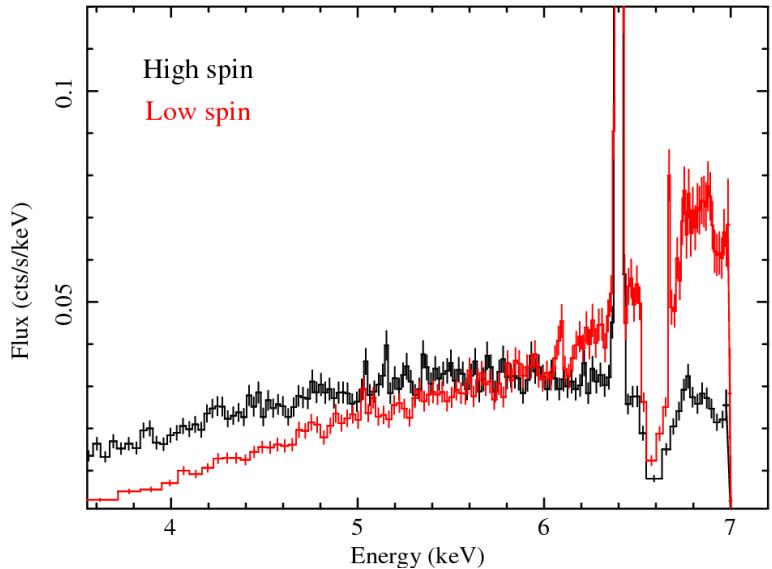

Figure 4: Simulated *Athena+*/X-IFU 150 ks iron line profile corresponding to a low spin (a=0) and high spin (a=0.998) black holes, superposed on the narrow emission feature at 6.4 keV emitted from a distant reflector (such as a molecular torus), plus ionized absorption from a wind ($Fe_{xxv}$ at ~ 6.6 keV). The flux of the source is ~ $10^{-11}$ cgs, typical of a moderately bright AGN. The disk inclination is 40º, the equivalent widths of the lines are 100 and 200 eV for the broad and narrow components, respectively, and a moderately thick and highly ionized wind were used for the simulations. *Athena+*/X-IFU will easily separate any narrow features from those produced by strong gravity.

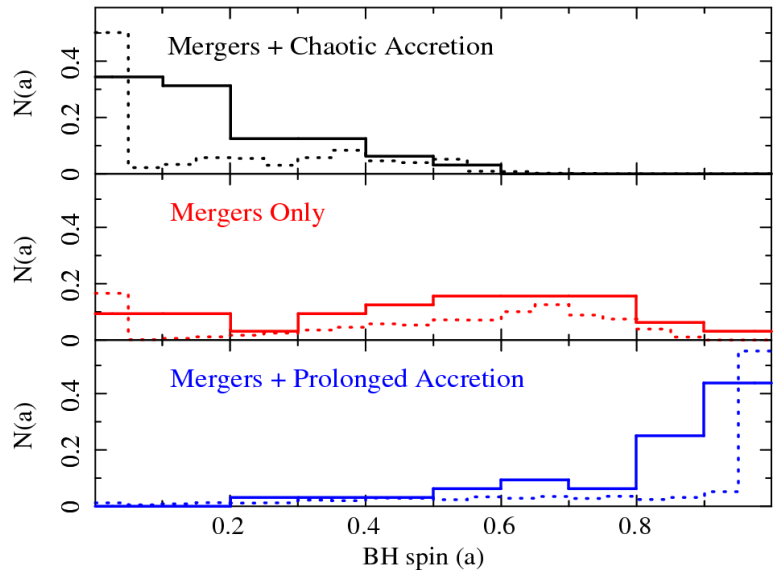

Figure 5: Spin as a probe of supermassive black hole growth history. The distribution of black hole spins in the local Universe depends on whether they have accumulated their mass predominantly via mergers, steady accretion or chaotic accretion. The theoretical expectations for each SMBH growth scenario (dotted histograms) is shown (Berti & Volonteri 2008) and compared to simulated *Athena+* measurements (solid histograms), accounting realistically for all observational errors and spectral complexities. The plot is made in the assumption that 50% of the brightest Seyfert 1 galaxies in the sky have a reflection component relativistically distorted (De la Calle Perez et al. 2010). Mean exposure time per source is 100 ks.

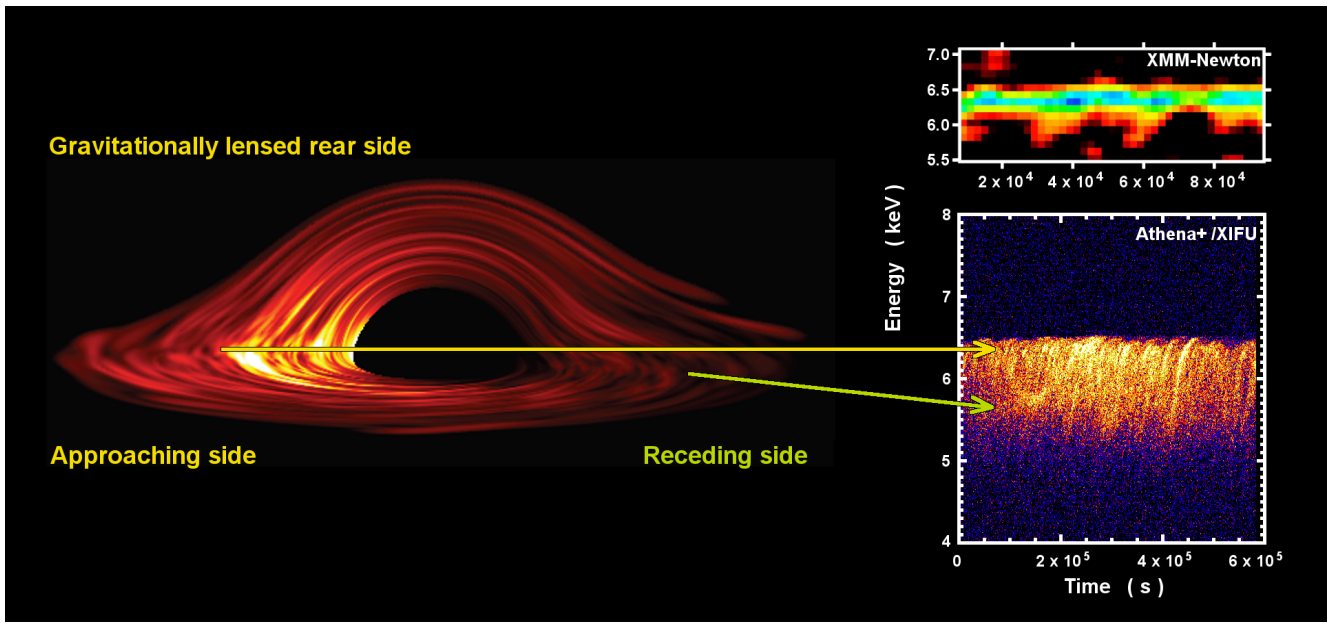

Figure 6: *Left:* Snapshot from a time-dependent Magneto-Hydrodynamic (MHD) simulation of an accretion disk around a BH (Armitage & Reynolds 2003). Rings and hotspots of emission are seen due to turbulence, the emission from which should be modulated on the orbital timescale. *Right:* Because the features are variable in both flux and energy, the disk can be mapped out in the time-energy plane. The first hints of this behaviour have been seen in XMM-Newton data (Iwasawa et al 2004, upper panel), but the improvement in throughput and especially in energy resolution offered by *Athena+* are needed to sample weaker and narrower features on suborbital timescales, allowing us to map out the inner accretion flow. In the lower panel the simulation with *Athena+*/X-IFU is shown, assuming a black hole mass of $3 \times 10^7$ solar masses, a 2–10 keV flux of $5 \times 10^{-11}$ cgs and a disk inclination of 20°.

*Athena+* will also be able to map the inner regions of the accretion disks in the time-energy plane. Any deviation from axial symmetry in the disk emissivity (e.g. associated with turbulence and/or the formation of hot spots) will lead to a characteristic variability of the iron line (Dovciak et al. 2004), with "arcs" being traced out on the time-energy plane (Armitage & Reynolds, see Fig. 6). Evidence of hot spots is found in XMM-Newton data (Iwawasa et al. 2004, De Marco et al. 2009; see also Reis et al. 2012 for a tidal disruption event in a dormant SMBH resulting in a quasi-periodic emission), and they are of great diagnostic power because they can be used to trace out the inner turbulent flow of the disk in the strong gravity environment. GR makes specific predictions for the form of these arcs, and the ensemble of arcs can in principle be fitted for the mass and spin of the BH, as well as for the inclination at which the accretion disk is being viewed, even if non-gravitational forces such as magnetic fields can also play a significant role in determining the hot spot evolution. *Athena+*'s large effective area and energy resolution at 6 keV are crucial for enabling the detection of these faint and relatively narrow structures, and for tracking them (in energy and flux) on the time scale of the innermost disk orbital periods, yielding new information on the geometry and physics of this region.

On a more speculative level, one can argue that the "arcs" described above can also be implicitly used as a probe and test of General Relativity in the strong field limit. More direct testing would involve comparing the observations with the predictions of different gravity theories. For instance, in the so-called pseudo-complex theory the line emission from an orbiting spot should have different timing and spectral characteristics due to the different values of the gravitational redshift and Keplerian frequency (Boller & Müller 2013). Subtle differences in the line profile are also expected if the no-hair theorem is violated (Johannsen & Psaltis 2012; according to the no-hair theorem, astrophysical black holes are fully characterized by their masses and spins and are described by the Kerr metric). While these kinds of measurements will certainly be very challenging, and their feasibility still needs to be fully addressed, their potential importance is such that they certainly deserve to be taken into consideration, especially for rapidly rotating black holes where relatively tight constraints on potential deviations from the Kerr metric are expected.

## 3. THE NATURE OF THE SOFT X-RAY EMISSION

In almost all radio-quiet AGN there is excess emission below 0.5-1 keV above that expected from an extrapolation of the higher energy power-law spectrum. As explained above, spectral and timing analyses indicate that, in many sources, reflection (both continuum and line emission) from an ionized accretion disk, blurred by strong gravity effects (e.g. Ross & Fabian 2005), provides an explanation for at least a part of this soft X-ray excess. However, in other objects there is evidence of a connection of the soft X-ray emission to the UV rather than to the hard X-rays (Mehdipour et al 2011), pointing to the existence of an additional Comptonised component which contributes to the soft X-ray excess (Done et al. 2012). There are sources in which there is no evidence for the strong relativistic blurring of the iron line which should be associated with a blurred reflection origin for the soft X-ray excess (e.g. Brenneman et al. 2012), even if extreme ionization of the disk cannot usually be ruled out. In fact, current data cannot distinguish spectrally between the additional Comptonised component and extremely blurred multiple reflectors, but *Athena+* with its large effective area in soft X-rays will easily distinguish between competing models, as illustrated in Fig. 7.

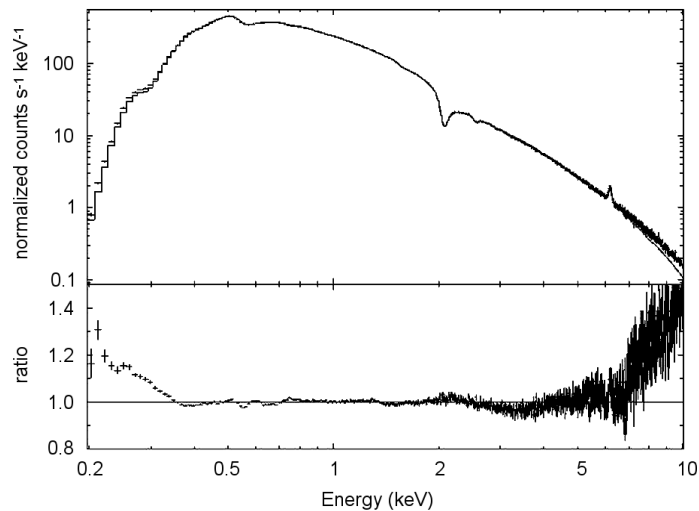

Figure 7: Simulated *Athena+* 50 ks spectrum with a Comptonization emission model for the soft X-ray excess (plus a hard power law and a cold reflection component) fitted with ionized, relativistically blurred disk reflection components. Parameters are those of the Seyfert 1 galaxy Ark 120, which is particularly suited for this purpose because of its negligible warm absorption (Vaughan et al. 2004). The two scenarios can be easily distinguished (data/model ratio shown in the bottom panel).

# 4. MAPPING THE CIRCUMNUCLEAR MATTER

The close environment of the black hole, defined as the sphere of influence (i.e. the region in which the gravitational potential of the black hole exceeds that of the host galaxy) is filled with a large amount of matter in different dynamical and ionization states. Using emission and absorption features produced by the circumnuclear medium, combined with techniques routinely used in optical astronomy, we can in principle measure the size and velocity structures of the emitting regions. A narrow component of the iron K$\alpha$ line is almost invariably present in the X-ray spectra of AGN (e.g. Yaqoob & Padmanabhan 2004, Nandra 2006). This line originates from reflection of the primary component by the circumnuclear gas; hence it is a powerful tool to establish the location of this gas (either the Broad Line Region, a parsec-scale torus, or the Narrow Line Region, the latter already outside the sphere of influence), which is related to the line width. In current observations the line is typically unresolved, with FWHM upper limits of several thousand km/s, compatible with any of the possible locations.

The unprecedented spectral resolution of the *Athena+*/X-IFU, combined with its large collecting area at 6 keV, will allow us to easily measure the FWHM and bulk velocity flow of any of the emitting regions using the neutral iron K$\alpha$ line in tens of sources (Fig. 8), thereby mapping the location and the relative amount of the different components. This technique will be especially useful in obscured AGN, where the continuum and inner disk emission are not detectable.

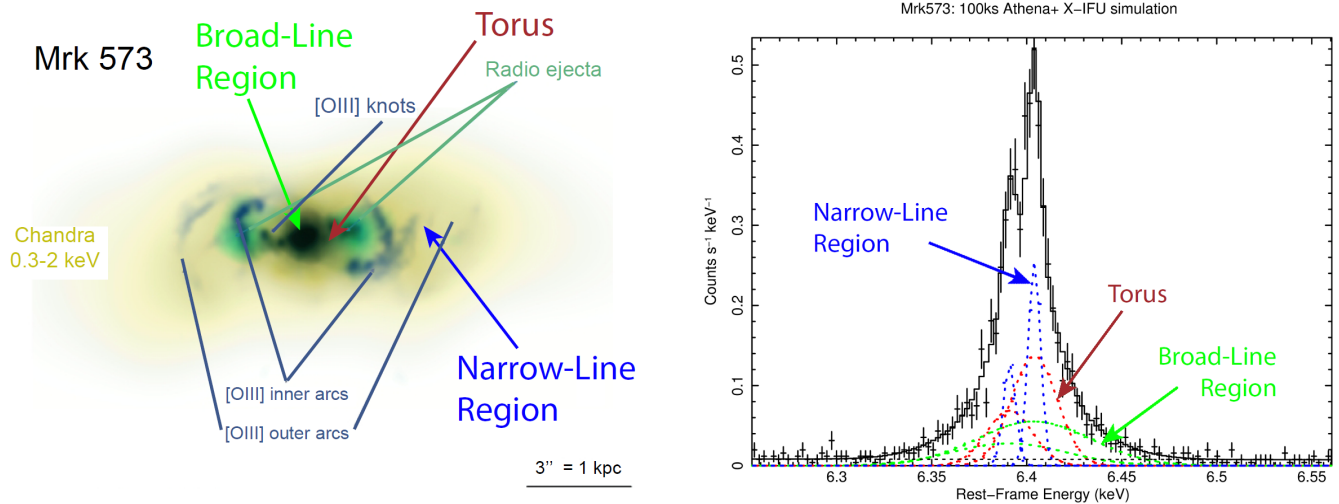

Figure 8: *Left:* The Seyfert 2 Mrk 573 multi-wavelength image combining soft X-ray (Chandra: yellow), radio 6 cm (VLA: green), and [Oiii] emission (HST: dark blue). The approximate locations of the main line-emitting regions are shown (some of them are not actually resolved in this image). *Right:* Simulated 200 ks *Athena+*/X-IFU observation of the neutral iron K$\alpha$ line (actually a doublet) arising from material at different distances from the black hole. The three components can be easily de-convolved if present in a total profile.